\documentclass[twocolumn,showpacs,amssymb]{revtex4}
\usepackage{graphicx}
\usepackage{dcolumn}
\usepackage{amsmath}
\usepackage{amssymb}
\usepackage{booktabs}
\begin{document}

\title{Statistically determined dispersion relations of magnetic field fluctuations in the terrestrial foreshock}

\author{B. Hnat$^{1}$, D. OConnell$^{1}$, V. M. Nakariakov$^{1}$}

\affiliation{$^{1}$ Centre for Fusion, Space and Astrophysics, University of Warwick.}
\date{\today}

\begin{abstract}
We obtain dispersion relations of magnetic field fluctuations for two crossings of the terrestrial foreshock
by Cluster spacecraft. These crossings cover plasma conditions which differ significantly in their plasma $\beta$
but not in the properties of the encountered ion population, both showing shell-like distribution function.
Dispersion relations are reconstructed using two-point instantaneous wave number estimations from pairs of Cluster
spacecraft. The accessible range of wave vectors, limited by the available spacecraft separations, extends to
$\approx \!2 \times 10^4$ km. Results show multiple branches of dispersion relation, associated with different power of
magnetic field fluctuations. We find that Sunward propagating fast magnetosonic waves and beam resonant modes are
dominant for the high plasma $\beta$ interval, while the dispersions of magnetically dominated interval include
Alfv\'{e}n and fast magnetosonic modes propagating Sunward and unit-Sunward.\\
\end{abstract}
\maketitle 

\section{Introduction.} The region upstream of the terrestrial bow shock provides a testbed for a large number of
generic plasma processes, from Fermi acceleration \cite{Bell78} and shock reformation \cite{Leroy83} to a wide range
of instabilities \cite{Quest88}. Due to the collisionless nature of the solar wind plasma, the kinetic energy of the
upstream flow is converted into heat by complex interactions of particles with fluctuations and through particle
reflection (acceleration) at the shock \cite{Fairfield69,Gosling82}. This leads to non-thermal velocity distribution
functions upstream of the shock and these can become linearly unstable to small perturbations, resulting in large
amplitude fluctuations in the magnetic field and other plasma parameters. Scattering and heating of ions by such
fluctuations is of great importance to understanding the channels of momentum and energy exchange in the system.
It is essential to keep in mind that the upstream flow is not laminar and that large amplitude, nearly harmonic
fluctuations coexist with turbulent solar wind. Identification of frequencies where the turbulent energy transfer
becomes dominant is of great importance to the determination of wave-particle interaction processes.

The unperturbed solar wind plasma near the Earth's orbit shows correlations between velocity and magnetic field
fluctuations, characteristic of transverse Alfv\'{e}n mode \cite{Bruno05}. The phase coherence of these fluctuations
is usually destroyed prior to their arrival at the foreshock, and only a single observation of pure Alfv\'{e}nic mode at
1AU has been reported \cite{Wang12} to date. Small density perturbations are also detected in the unperturbed solar
wind and these are a mixture of the pressure-balnced structures of coronal origin and fast magnetosonic modes \cite{Tu94}.
The ion cyclotron waves, Alfv\'{e}n waves with frequency near the ion cyclotron frequency and with the field-aligned wave
vector, have been suggested as a possible source of solar wind heating and acceleration, but unambiguous detection of
these waves in the solar wind has been elusive, see for example \cite{Kasper13}. Upon entering the foreshock, the solar
wind plasma interacts with the small population of reflected particles and this interaction modifies kinetic
properties of the plasma and can lead to a significantly different dynamics.

Few classes of ion velocity distribution functions have been observed in the terrestrial foreshock \cite{Gosling78} and,
in the case of the quasi-parallel ($\Theta_{B,n} < 45^{\circ}$ bow shock, these distributions have been associated with
low frequency electromagnetic fluctuations. Fast magnetodonic waves, travelling in the direction of the beam, are the
fastest grown unstable modes destabilised by the cold and tenuous beam-like distributions, corresponding to ions travelling
Sunward and nearly parallel to the background magnetic field. For sufficiently dense and warm beam the destabilised fast
wave can also propagate agains the beam \cite{Gary85}.  In addition to the beam-like distributions there are also these of
``diffuse" type \cite{Gosling78,Paschmann81}, which are nearly isotropic in the phase space, and have large temperatures,
sometimes exceeding $10$ keV. Enhanced wave activity has been observed for these distributions, with both left and
right-hand modes competing in growth rates, in numerical simulations with different beam and plasma parameters
\cite{Sentman81}. We note the importance of minor ions, especially Helium He$^{++}$, which may resonate with the left-hand
mode at low frequencies, where the power of fluctuations tends to be greater and provide energy source for magneto acoustic
cyclotron instability \cite{Dendy94}. 
\begin{table}[h!]
  \centering
  \caption{Cluster intervals parameters.}
  \begin{tabular}{ccccccccc}
\toprule
Interval & $\chi$[km] & {\bf B}[nT, GSE] & $\langle V_{sw} \rangle$ & $\beta$ & $V_A$ & $\Theta_{V,B}$ & $\omega_{cp}$ & $\rho_i$[km]\\
\hline
I1 & 70.5 & (-6.9, 0.3, 1.3) & 315.5 & 0.4 & 60.5 & 21 & 0.7 & 42.1\\
I2 & 82.3 & (7.9, -3.9, 4.7) & 406.6 & 3.2 & 93.0 & 147 & 1.0 & 177.5\\
\bottomrule
\end{tabular}
\label{table:dates}
\end{table}
In principle, the range of frequencies supported by foreshock plasmas span many orders of magnitude, from ultra-low frequency
(ULF) waves at about $10^{-2}$ Hz to electron Langmuir waves at hundreds of kHz. A substantial experimental literature exists,
where the dominant modes at low frequencies have been characterised in terms of their wave properties in different regions of
the terrestrial foreshock \cite{HoppeRussell83,Le91,Fazakerley95,Eastwood03}. The focus on ULF waves is motivated by numerical
simulations \cite{Gary98,Sentman81,Winske84}, which established that, for realistic plasma parameters,
beam-like distributions generate electromagnetic fluctuations in the low frequency range, that is, below the ion cyclotron
frequency. More recent work took into account other plasma parameters, and their correlations with the magnetic field fluctuations,
providing more unambiguous identification of magntohydrodynamic (MHD) modes (e.g., \cite{Blanco-Cano97}).
\begin{figure} [ht]
\includegraphics[width=1\columnwidth]{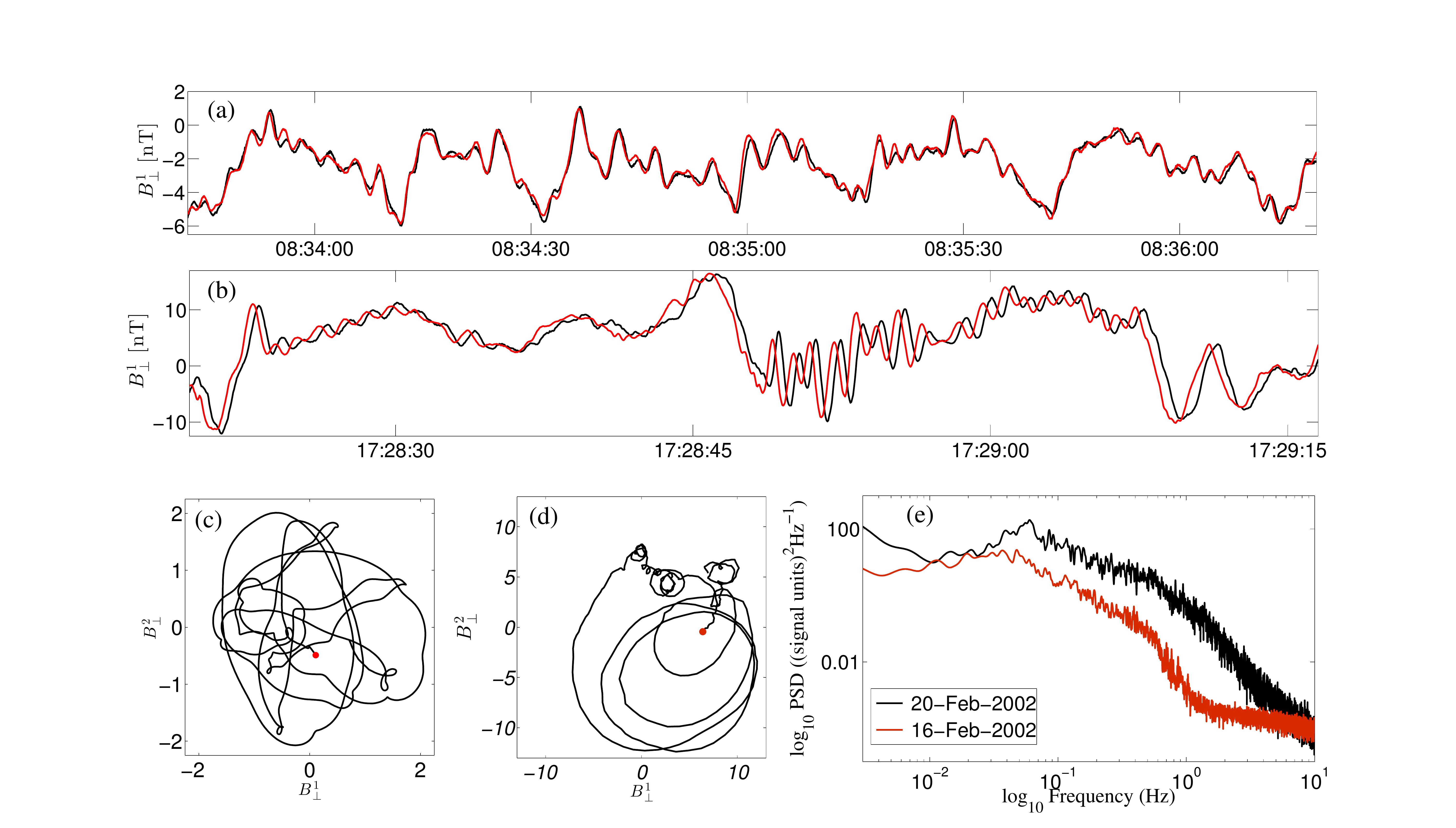}
\caption{Magnetic field data summary for two time intervals. Panel (a): transverse magnetic field component associated with
the largest eigenvalue of the minimum variance analysis for a section within $I1$ and for spacecraft $C3$ (black trace)
and $C4$ (red trace). Panel (b): same as (a) for the interval $I2$. Panels (c,d): hodograms of the transverse magnetic
field components for selected short subintervals in $I1$ and $I2$, respectively. Panel (e) power spectral density for
$B_{\perp}^1$ component of interval $I1$ (red) and interval $I2$ (black).}
\label{fig1} 
\end{figure}

Multi-point Cluster observations, aided with the wave telescope technique, provided estimates of the dispersion relations
associated with the dominant modes in the low frequency range and clearly demonstrated the existence of Sunward propagating
waves, broadly consistent with the ion beam instability as a generation mechanism \cite{Narita03}. 
Given the multiple locations where unstable distributions may exist, the growth rates of relevant instabilities
(order 100 seconds) and the speed of the propagation (typically 50-100 km/s) one expects a continues distribution
of power for unstable modes \cite{Sentman81}. In addition, nonlinear processes, for example, decay instability of
Alfv\'{e}n waves or/and the modulation instability of the fast magnetosonic mode \cite{Hollweg94}, may lead to the
re-distributed of power.

Our purpose in this work is to apply a new technique, which clearly identifies how fluctuating power is distributed
among different quasi-coherent modes in two foreshock plasmas, which differ in their macroscopic parameters, but have
similar ion distributions. This work extends previous multi-point studies by applying a statistical method \cite{Beall},
capable of detecting dispersion relations associated with different coexisting modes within the same set of fluctuations.
Each time interval is treated as a set of independent shorter realisations, each contributing to a statistically generated
dispersion relation represented as the cumulative power at each identified frequency and the corresponding wave number.
We find that, for anisotropic and warm beam, the most energetic fluctuations are located on at the right-hand polarised mode
and resonant ion beam branches of the dispersion relation. A smaller, but considerable, amount of power, is also found in
the left-hand polarised branch, which extends to large modules of wave number, $k \rho_p \sim 0.75$. For the interval with
diffuse-like ion distribution we find the dominant power at low frequencies to be associated with coherent structures,
which may correspond to turbulent vortices with slow dynamics, that is, dominated by advective solar wind transport.

\section{Data.} The dataset consists of two foreshock crossings, $16/02/2002$ at $07$:$50$-$09$:$20$ and $20/02/2002$ at
$16$:$56$-$17$:$56$, hereafter referred to as $I1$ and $I2$, respectively. These foreshock crossings were selected based
on the clear presence of wave modes in frequency spectra as well as for the small spacecraft separation of the spacecraft
pair $C3$ and $C4$. We use magnetic field data from the FGM instrument \cite{Balogh97} with sampling frequency of
$\approx 22.5$ Hz and spin resolution ($4$ seconds) CIS-HIA data \cite{Reme01} for plasma parameters.
Table \ref{table:dates} presents a summary of the plasma parameters for both intervals. We note a large difference in ion
plasma $\beta_i$ for these intervals as well as a different direction of the main component of the magnetic field (GSE $x$
direction). For the interval $I1$ the magnetic field is directed towards the Earth, while interval $I2$ has magnetic field
pointing Sunwards. Both intervals has been studied before in the context of ULF waves \cite{Narita04,Narita03,Lucek04}
and their possible impact on temperature anisotropy of the core ion population \cite{Selzer14}. 

Our analysis is undertaken for transverse fluctuations of the magnetic field expressed in minimum variance coordinates.
The ratio of the intermediate and minimum eigenvalues was $\sim 2.5$, which we contribute to the non-stationary character
of these considerably long intervals.
Figure \ref{fig1} presents a summary of the magnetic field data for the two intervals. These intervals show typical values
of the ion number density ($\langle n_p \rangle \sim 5-10$ cm$^{-3}$) of the solar wind and their bulk speed puts them in
the slow solar wind category. Clear wave trains are seen in panels (a,b), in which we plot snapshots of the transverse
magnetic field component corresponding to the highest eigenvalue of minimum variance analysis. Two traces show data from
spacecrafts $C3$ (black) and $C4$ (red). These demonstrate that the observed temporal shift between the signals is never
larger than a period of a wave. Panels (c,d) demonstrate that, for sufficiently short subintervals, a unique sense of circular
polarisation can be established in the spacecraft frame. While interval $I2$ shows fairly consistent left-hand polarisation,
interval $I1$ exhibits frequent changes from left to right-hand polarisation. The frequency spectra of transverse magnetic
field components, shown in panel (e), reveal peaks at frequencies lower then the proton gyro-frequency
($0.01 \leq \nu \leq 0.1$ Hz) and a broadband, power law behaviour at higher frequencies for both intervals. The proton
cyclotron frequency in the spacecraft frame has been estimated at $\nu_{cp} \approx 0.11$ Hz, but a large Doppler shift puts
it in the range of $3-5$ Hz in the plasma frame. Table \ref{table:dates}, gives values of gyro frequencies (angular frequencies,
in [rad/s]) in the plasma frame of reference.
\begin{figure*} [ht]
\centering
\includegraphics[width=0.4\textwidth]{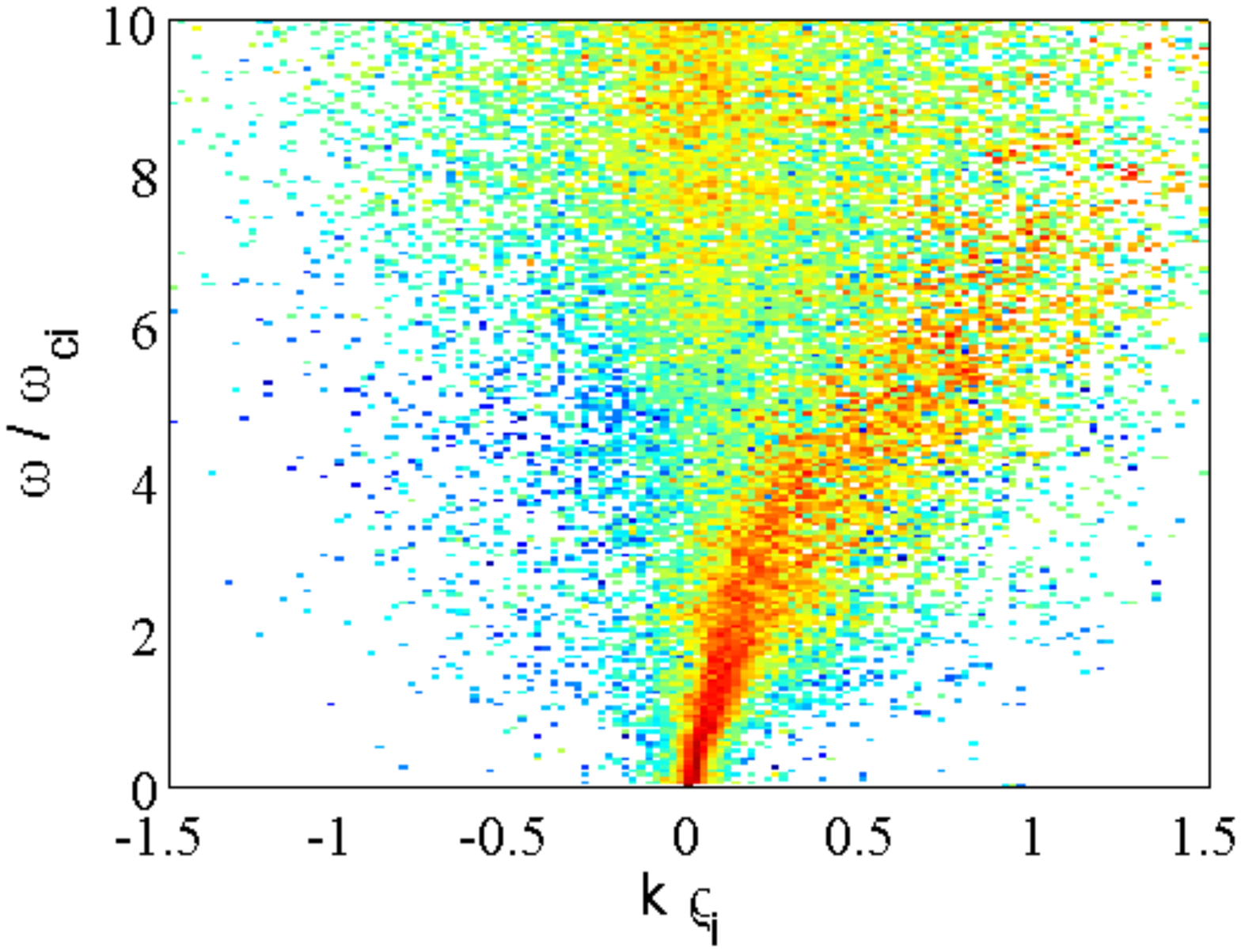}   
\includegraphics[width=0.4\textwidth]{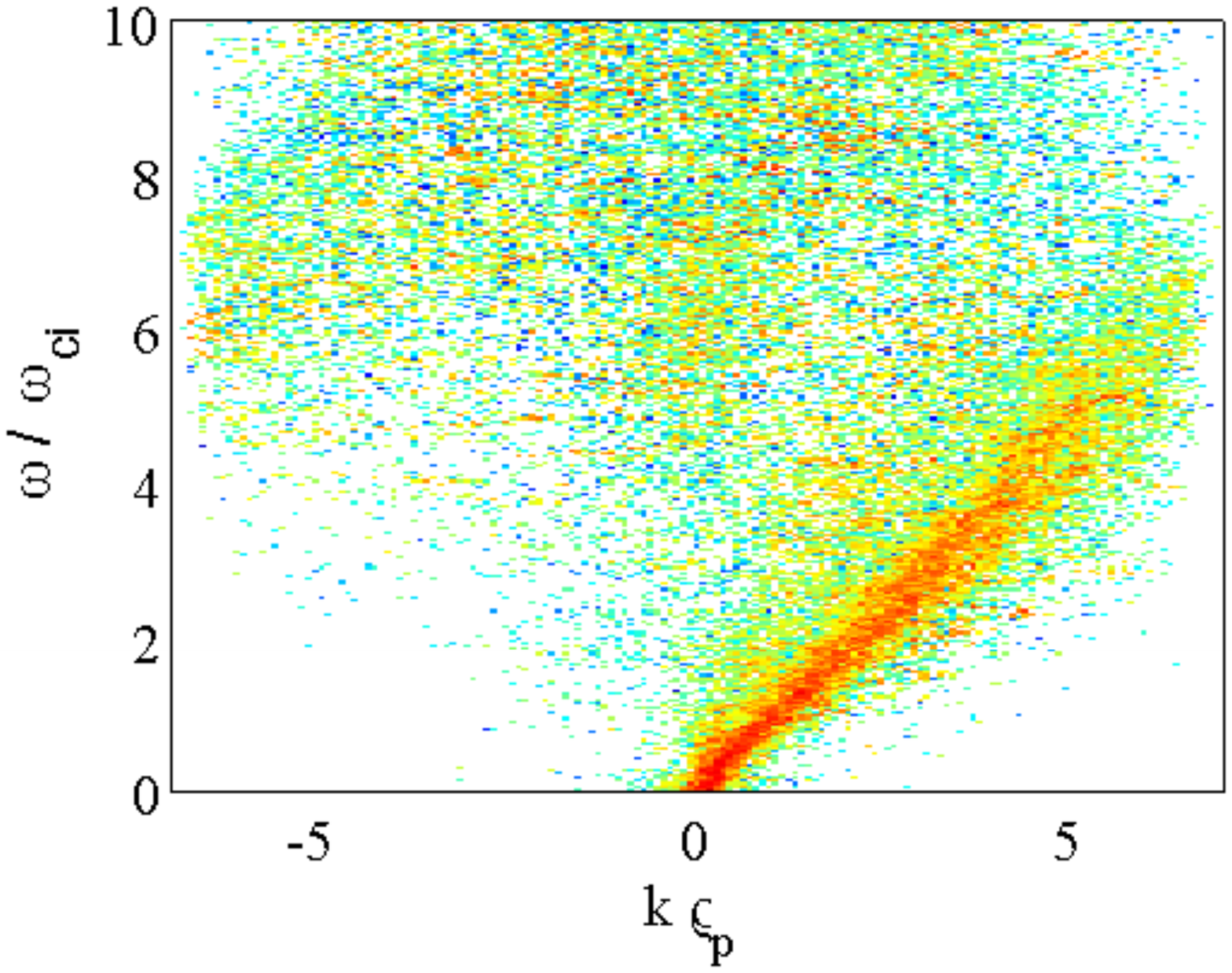}   
\caption{Histograms of normalised cumulative power on normalised frequency-wave number plane for each interval.}
\label{fig2ab}
\end{figure*}
\section{Methodology.} 
Transverse magnetic field components, from spacecrafts $C3$ and $C4$ are used to construct an estimate of the
dispersion relation. We use a Fourier technique based on Beall's algorithm \cite{Beall,Dudok95,Balikhin97}
designed for fixed probe pairs. Let $\bm{B}(\bm{x},t)$ represent a zero-mean, stationary, homogeneous vector field,
which can be represented as a superposition of plane waves, that is,
\begin{equation}
\label{eq:Btilda}
\bm{B}(\bm{x},t)=\int_{-\infty}^{\infty} d\bm{k} \int_{-\infty}^{\infty} \tilde{\bm{B}}(\bm{k},\omega) \exp[i(\bm{k}\cdot \bm{x}-\omega t)],
\end{equation}
where the integrals are taken as the discrete sum of the measurements.
The goal is to calculate an estimate of the dispersion relation, $\bm{k}=\bm{k}(\omega)$, which in the case of two probes at
$\bm{x}_1$ and $\bm{x}_2$, is given by a projection $k_{\chi} = \bm{k}(\omega) \cdot \bm{\chi}$, where $\bm{\chi}=\bm{x}_2-\bm{x}_1$.
Essentially, all required information is contained in the cross-spectral density, $H(\bm{\chi},\omega)$, which is defined as:
\begin{equation}
\label{eq:H}
H(\bm{\chi}, \omega) = \lim_{T\to\infty} \frac{1}{T} \langle \tilde{\bm{B}}^{*}(\bm{x}_1,\omega) \tilde{\bm{B}}(\bm{x}_2,\omega)\rangle,
\end{equation}
where $T$ is the total time of the measurements and the angular brackets represent time averages. Expressing the complex quantity
$H(\bm{\chi}, \omega)$ in polar representation, allows the definition of the local wave number, $k_{\chi}$ in terms of the phase:
$\Theta(\bm{\chi}, \omega)=Im(H(\bm{\chi}, \omega)) / Re(H(\bm{\chi}, \omega))$,
\begin{equation}
\label{eq:K}
k_{\chi}(\omega)=\frac{d}{d\bm{\chi}}\Theta(\bm{\chi},\omega) \approx \frac{\Theta(\omega)}{\chi}.
\end{equation}

In order to resolve a dispersion relation of multiple co-existing branches, it is necessary to calculate $k_{\chi}$ for an ensemble of
data sets. Accordingly, for each interval in table \ref{table:dates}, the data is partitioned into $N$ overlapping
realisations of length $T$, chosen to include time scales of interest. The quantities given in (\ref{eq:H}) and (\ref{eq:K})
are then calculated for each realisation. This yields the local joint wavenumber-frequency spectrum
\begin{equation}
\label{eq:S_hat}
\hat{S}(k_{\chi},\omega)=\langle \frac{1}{2}(|\tilde{\bm{B}}(\bm{x}_1,\omega)|^2+|\tilde{\bm{B}}(\bm{x}_2,\omega)|^2)\delta(k-k_{\chi}) \rangle.
\end{equation}
where angular brackets now indicate an ensemble average. A histogram of discrete cells in $k$ and $\omega$ for
$\hat{S}(k_{\chi},\omega)$ is constructed to give the dispersion relation. Equation (\ref{eq:S_hat}) converges to a real
dispersion relation provided the plane wave approximation (\ref{eq:Btilda}) is valid and that $\bm{k} \cdot \bm{\chi}<2 \pi$. 
The quantity $\hat{S}(k_{\chi},\omega)$ is then a discrete two-dimensional histogram of the cumulative power on a
$(k_{\chi},\omega)$ plane in the spacecraft reference frame.

In order to identify dispersion relations in the plasma rest frame, we apply the Doppler shift to $(\omega, k_{\chi})$ pairs
which correspond to the maxima in $\hat{S}(\omega)$ at each frequency. In practice, we find three highest values of power,
because some peaks may be over-resolved for a given number of bins in wave number and give the same Doppler shifted dispersion
as the first maximum, for some frequencies. The procedure is then as follows: we identify three highest values of power in
$\hat{S}(\omega)$ at each frequency, associate each frequency with its corresponding wave number $k_{\chi}$ and Doppler shift
these using the mean solar wind velocity, that is:
\begin{equation}
\label{eq:Dopler}
\omega_{pl}=\omega_{sc} - k_{\chi} \langle V_{sw} \rangle \cos(\Theta_{\bm{v},\bm{\chi}}).
\end{equation}
We have chosen the convention whereby $\omega$ is always positive and we therefore switch the sign of the wave number
$k_{\chi} \to -k_{\chi}$ when Doppler shifted frequency $\omega_{pl} < 0$.
Finally, we choose a sign of $k_{\chi}$ in GSE coordinate system, that is, $k_{\chi} > 0$ for waves travelling Sunwards
and $k_{\chi} < 0$ for waves travelling towards the bow shock in the plasma rest frame.
\begin{figure*} [ht]
\centering
\includegraphics[width=0.45\textwidth]{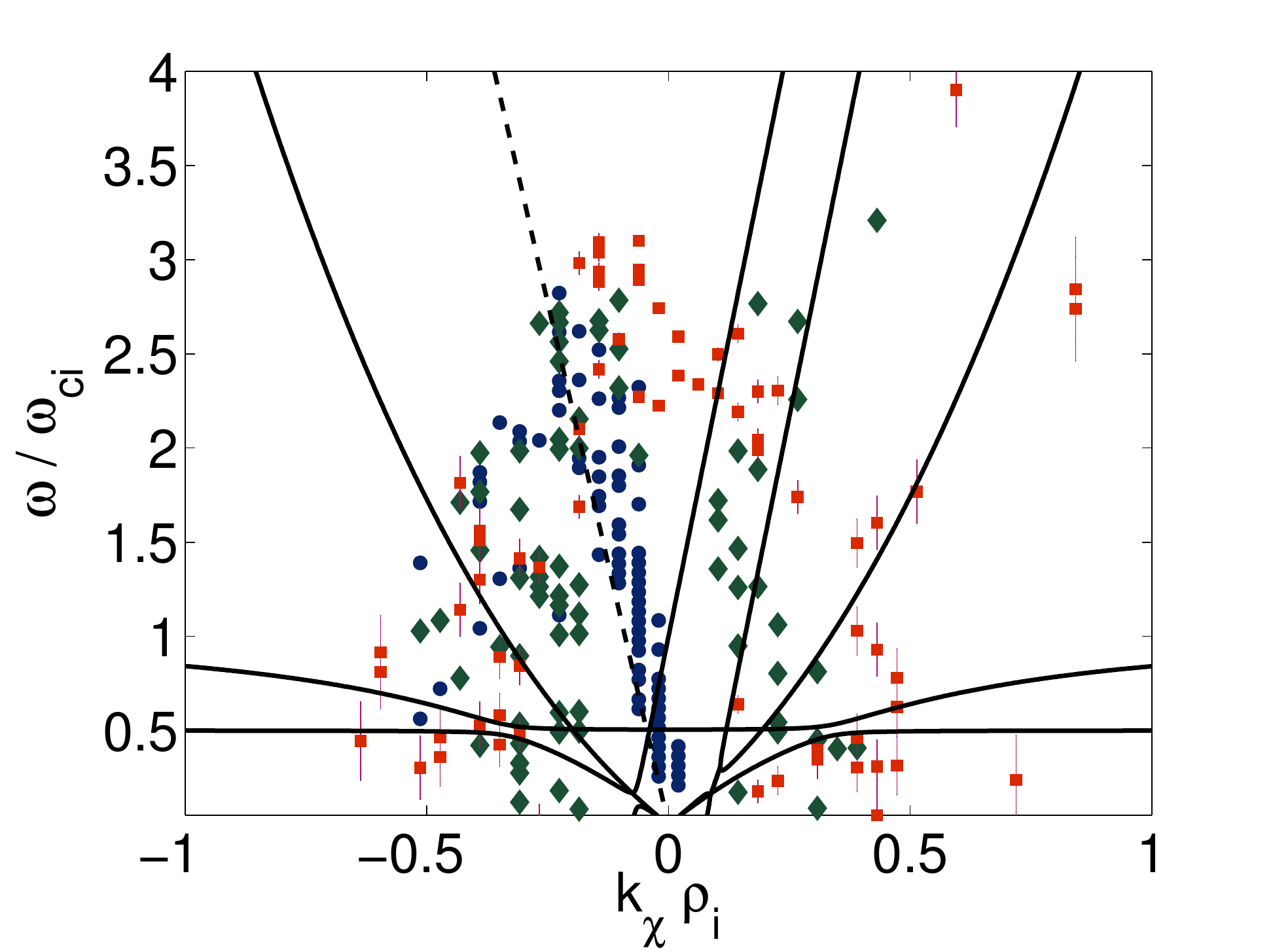}   
\includegraphics[width=0.45\textwidth]{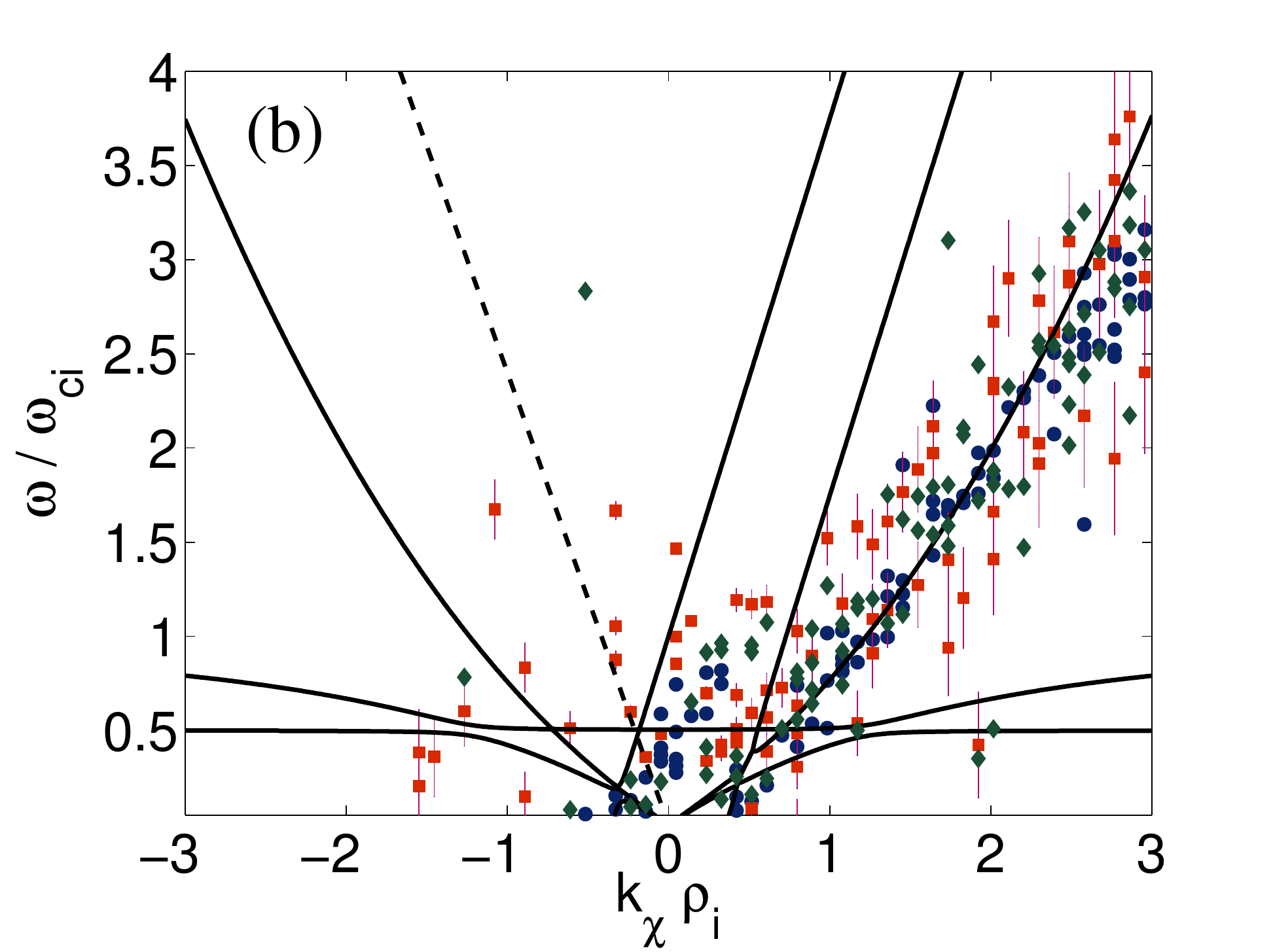}   
\caption{Doppler shifted estimates of dispersion relations obtained for each interval (solid points) and cold
plasma dispersion relations as a function of wave vector amplitude (solid lines). Symbols correspond to
different maxima in fluctuation power in $\hat{S}$ (see figure \ref{fig2ab}): the first maximum (highest power for a given
frequency) is shown as circles, the second maximum is shown as diamonds and the third as squares. In panel (b) diamond
symbols have been removed for clarity, they follow the same pattern as filled circles.
Errors are from the Doppler shift uncertainty.}
\label{fig3ab}
\end{figure*}
\section{Results and Discussion} The two panels of Figure \ref{fig2ab} show images of the cumulative power,
$\hat{S}$, on the $(\omega_{sc}/\omega_{cp}, k_{\chi} \rho_i)$ plane, where $\rho_i$ is the ion gyro-radius
calculated using the CIS-HIA measured perpendicular temperature (with respect to magnetic field) and proton mass.
The Nyquist frequency of the magnetic field measurements is $12.5$ Hz, but we only show frequencies where
some correlation of $\omega_{sc}$ and $k_{\chi}$ clearly exists. Beyond this range turbulent fluctuations
dominate, producing a random, broad signature, without any apparent features. Wave number space in the range
allowed by the method, that is $(-\pi / \chi,+\pi / \chi)$, which in SI units corresponds to wave length between
$(-2.5, 2.5) \times 10^4$ meters, has been divided into $130$ equally spaced bins. We note that the colour
scale used in the figure is that of $\log_{10}(\hat{S})$. These plots were generated using $\sim\!100$ strongly
overlapping ($\sim\!75\%$ overlap) realisations, each $5$ minutes long. The data for each realisation has
been detrended and we subtracted the mean prior to obtaining the Fourier components. The figure demonstrates
that a clear wave activity can be found when applying the method outlined above, but strong Doppler shift
obscures any details of the dispersion relations.

As outlined before, we find the three highest local maxima in $\hat{S}$ and apply the Doppler shift to each
identified $(\omega_{sc}/\omega_cp, k_{\chi} \rho_i)$ pair, as given in (\ref{eq:Dopler}). These dispersion
relations, in normalised quantities, are shown in figure \ref{fig3ab} for each interval. Solid lines represent
the cold plasma dispersion relations with massless electrons, background protons, background He$^{++}$ ions and
a proton beam \cite{Verscharen13}. We have used the wave propagation angle (w.r.t magnetic field vector) of
$10^{\circ}$ and beam velocity of $4.25 V_A$ for $I1$ and the angle of $30^{\circ}$ and $5 V_A$ for $I2$,
where $V_A$ is the average Alfv\'{e}n speed. The beam velocity and the wave propagation angle were not derived
from the data, but were chosen based on the visual agreement of the theoretical dispersion curves with experimental
observations. In this context, the theoretical curves are given to guide the eye, but should not be compared
directly to the discrete points, since $k_{\chi}$ is a projection of the estimated wave vector on the spacecraft
separation vector and its true angle to the magnetic field is not known. Different symbols used in the figure
correspond to the power associated with the local maxima in $\hat{S}$, filled circles represent the highest
power, filled diamonds and squares correspond to the second and the third peak, respectively. The errors shown
are due to the uncertainty in the Doppler shift with the turbulent velocity field and have been calculated based
on the variance of the solar wind velocity.

There is a clear difference between the two intervals considered here. For the interval $I1$, the most powerful
fluctuations (filled circles) are convected structures, for which Taylor hypothesis is approximately satisfied,
that is $\omega_{pl} \approx k_{\chi} V_{sw}$ (dased line). A small fraction of the most powerful fluctuations
is also positioned near the fast magnetosonic and Alfv\'{e}n branch of the dispersion curves, for $k_{\chi}<0$.
These are waves travelling towards the bow shock and are most likely associated with he solar wind fluctuations.
The points obtained from the secondary and the tertiary peak in power, shown as diamonds and squares, respectively,
are equally distributed between fast magnetosonic and Alfv\'{e}n modes propagating towards and away from the bow
shock as well as the beam resonant mode. We have measured the ratio of power for the two dominant peaks,
as a function of frequency, $PR(\omega)=\hat{S}(\omega) / \hat{S}_{max}(\omega)$. We found that this ratio shows
a large increase around $\omega \approx 0.5 \omega_{cp}$, so that $PR(\omega<0.5 \omega_{cp})=0.02$, while
$PR(\omega > \omega_{cp})=0.6$. The largest $k_{\chi}$ value obtained for the Sunward propagating Alfv\'{e}n branch
($k_{\chi}>0$) is approximately $k_{\chi} \rho_p=0.5$, which is much larger than the numerically found value for the
most unstable left and right-hand polarised mode driven by cold beam \cite{Gary85}. Finally, we note a suggestive
the presence of points near the ion cyclotron branch of helium, $He^{++}$ for $I1$. This branch is difficult to study,
since the lower frequency modes require longer realisations to be considered. However, for the plasma that supports
Bernstein modes, the minor ion resonance can contribute to magnetoacoustic cyclotron instability \cite{Dendy94}.

We contrast this with the finding from interval $I2$, where all three most powerful modes are predominantly these of
the right-hand polarised fast magnetosonic type, propagating Sunwards. We note the presence of points along the straight
solid line for $\omega<0$ and $k_{\chi}<0$, which coincide with the low frequency proton beam resonant modes.
Similar to the previous interval, the secondary and tertiary populations, appears to follow the $He^{++}$ cyclotron
branch for modes propagating Sunwards, some of them have $k_{\chi} \rho_p \ll 1$, which may indicate a large
perpendicular wave number component. Finally, some residual amount of power is found in the modes propagating towards
the bow shock. This result while consistent with the previous analysis of this interval \cite{Narita03}, supports
numerical results which indicate that the unstable modes associated with the ``diffused" ion population are both
left and right-hand polarised modes propagating Sunward. We attribute residual power in the anti-Sunward propagating
modes with the solar wind population. 
\begin{figure} [h]
\includegraphics[width=1\columnwidth]{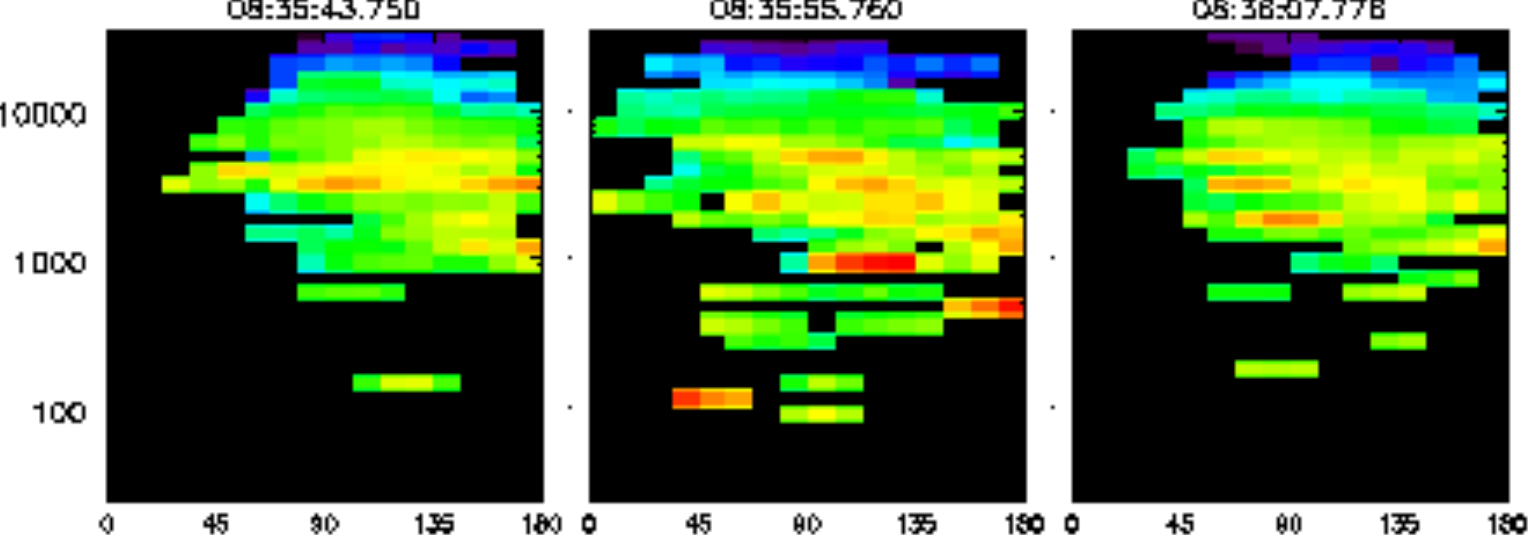}
\includegraphics[width=1\columnwidth]{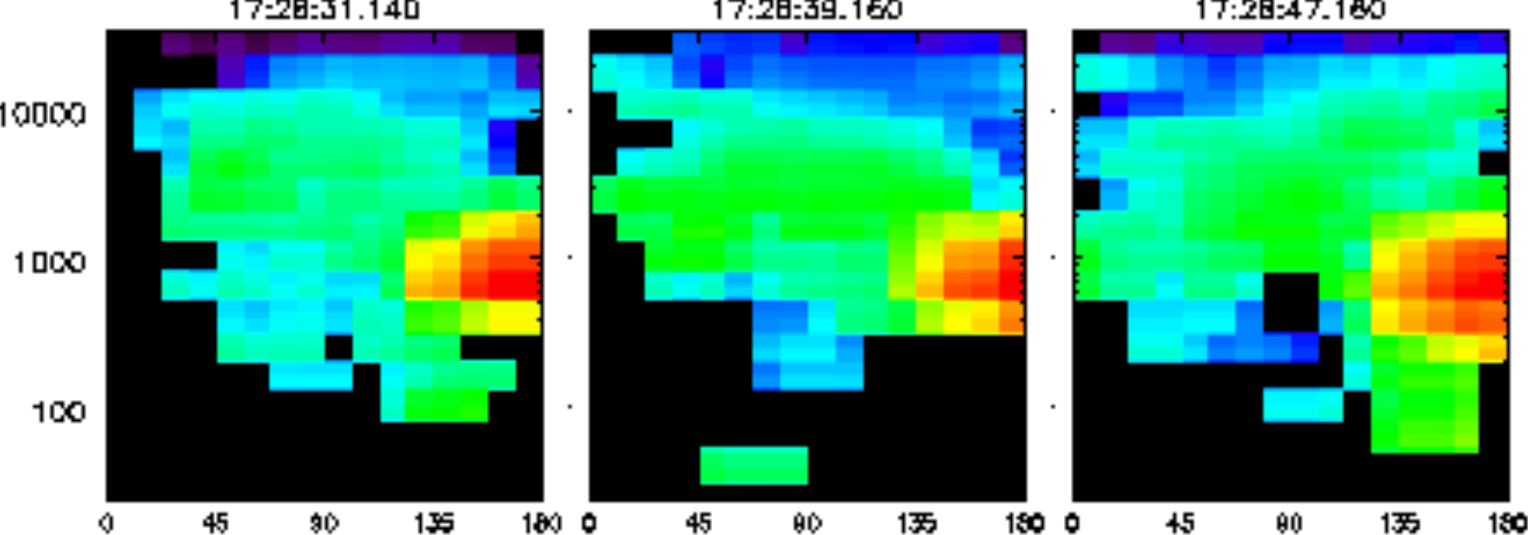}
\caption{Differential proton flux (colour scale), calculated from high resolution HIA proton spectra, for the total energy
as a function of pitch angle for approximately $1$ minute period within each studied interval. Horizontal axis gives pitch
angle in degrees, vertical one shows the energy in eV.}
\label{fig4ab} 
\end{figure}

Having established signatures of multiple dispersion relations in each interval we now examine proton distribution functions.
Figure \ref{fig4ab} shows CODIF proton pitch angle versus energy particle flux distributions, observed within $\approx 25$
seconds window in intervals $I1$ (top panel) and $I2$ (bottom panel), respectively. While this observation window is very
short compared with the total length of data, we have checked by visual inspection that these distributions provide a good representation of the ion population observed throughout each interval. CIS instruments were in the solar wind mode for interval
I1, and in the magnetospheric mode during the second interval, I2. In the magnetospheric mode the instrument samples all angular
directions and it is clear that the solar wind beam dominates the spectrum in this case. Accounting for this operational effect,
both intervals show relatively typical shell-like ion distributions characteristic of the quasi-parallel foreshock. In both
cases the peak ion energies are few keV and the distributions have some angular asymmetries. the CIS-CODIF spectra show
low level of the solar wind flux contamination for the interval I1, with the warm proton beam at energy of about 10 keV.

\section{Conclusions}
We have identified multiple dispersion relations coexisting within the turbulent plasma for two different crossings
of the terrestrial foreshock by Cluster spacecraft. These dispersions span beyond typically studied ULF waves in
frequencies and wave numbers, with the maximum angular frequency exceeding the proton gyro-frequency by a factor
of up to four. Constructed dispersion relations differ strikingly for the two intervals we have studied. Interval I1,
while dominated by advocating structures, has significant amount of power in Alfv\'{e}n and fast magnetosonic waves
propagating in Sunward and unit-Sunward directions. Examined distribution functions show a typical ``diffused" ions
with energy around 10 keV. While for highly anisotropic beam temperatures ($T_{\perp,b} > T_{\parallel,b}$) the
instability can generate ion cyclotron waves propagating parallel and anti-parallel to the magnetic field \cite{Gary85},
we must also admit a possibility that the anti-Sunward propagating fluctuations are these embedded in the solar wind.
Importantly, the presence of counter-propagating Alfv\'{e}n modes may give rise to their nonlinear interaction.
We have already highlighted a possibility of ion cyclotron modes of $He^{++}$ and their importance for the
magnetoacoustic cyclotron instability.

Interval I2 shows the dispersion relation dominated almost entirely by fast magnetosonic and resonant proton beam
modes, indicating an instability of the cold ion beam, for which $\omega_{pl} = k_{\parallel} v_b - \omega_{cp}$,
where $v_b$ is the speed of the beam. While observed distribution functions for this interval show a typical
``diffused" population, the instrument mode of interval I2 is dominated by the solar wind beam and is difficult to
interpret. Previous study of this interval concluded that the observed fast magnetosonic waves are due to cold beam
instability, since there is evidence that the observed dispersion includes beam resonant modes \cite{Narita03}.
The right-hand polarised waves of the interval $I2$ reach speeds of up to $v_{ph}^R \approx 170$ km/s.

\textbf{Acknowledgments: } We acknowledge the CLUSTER team for data provision. This work was supported by the UK STFC
and EU Marie Curie ``Turboplasmas" funding.

\end{document}